\documentclass [11pt,twoside]{article}

\usepackage{shrthnds}
\usepackage{cite}
\usepackage{graphicx}
\usepackage{color}
\usepackage{subfigure}
\usepackage{setspace}

\usepackage{multirow}

\usepackage[hang,small]{caption}

\usepackage{geometry}
\usepackage{fancyhdr}
\usepackage{titlesec,titletoc}
  \titleformat{\section}{\Large\sf\bfseries}{\thesection}{1em}{}
  \titleformat{\subsection}{\large\sf\bfseries}{\thesubsection}{1em}{}

\title{\sf\bfseries 
Relating the cosmological constant and slow roll to conformal symmetry breaking}

\author{\normalsize  
Pankaj Jain\footnote{email: pkjain@iitk.ac.in}~,
 and Gopal Kashyap\footnote{email: gopal@iitk.ac.in}}
\date{}%{\today}

\begin{document}
\maketitle
\vspace{-0.6cm}
\bc
{\small Department of Physics, IIT Kanpur, Kanpur 208 016, India}
\ec

\centerline{\small\date{\today}}
\vspace{0.5cm}

\bc
\begin{minipage}{0.9\textwidth}\begin{spacing}{1}{\small {\bf Abstract:}
%%%%%%%%%%%%%%%%%%%%%%%%%%%%%%%%%%%%%%%%%%%%%%%%%%%%%%
% ABSTRACT
%%%%%%%%%%%%%%%%%%%%%%%%%%%%%%%%%%%%%%%%%%%%%%%%%%%%%%
We show that a theory with conformal invariance, which is explicitly broken
by small terms, provides a solution to the
fine tuning problem of the cosmological constant.
In the absence of the symmetry breaking terms, the cosmological constant
is zero. Its value in the full theory is controlled by the symmetry
breaking terms. The symmetry breaking terms also provide the slow 
roll conditions, which may be useful in constructing a model of inflation.

%%%%%%%%%%%%%%%%%%%%%%%%%%%%%%%%%%%%%%%%%%%%%%%%%%%%%%
}\end{spacing}\end{minipage}\ec

\vspace{0.5cm}\begin{spacing}{1.1}

%%%%%%%%%%%%%%%%%%%%%%%%%%%%%%%%%%%%%%%%%%%%%%%%%%%%%%
% MAIN CONTENT STARTS HERE
%%%%%%%%%%%%%%%%%%%%%%%%%%%%%%%%%%%%%%%%%%%%%%%%%%%%%%
\section{Introduction}

It has
been argued that if conformal invariance is broken by a soft mechanism then
it might be possible to preserve its consequences even in the full quantum
theory \cite{Englert:1976,Shaposhnikov:2008a,Shaposhnikov:2008b,JMS,JM10_2,JMPRD}. This allows the possibility of constructing a theory in
which the cosmological constant can be set identically to zero. 
Let us consider the mechanism proposed in \cite{Shaposhnikov:2008a}. 
We consider a toy model with two real scalar fields. 
The Lagrangian density, in four dimensions, may be written as, 
\begin{equation}
{\cal L} = {\cal L}_G + {\cal L}_M + {\cal L}_{SB}
\label{eq:Lagrangian}
\end{equation}
where ${\cal L}_G$ and  ${\cal L}_M$ are the gravity and matter Lagrangian
\begin{equation}
{\cal L}_G = - (\beta^2_1 \chi^2 + \beta^2_2\phi^2)R 
\end{equation}
\begin{equation}
{\cal L}_M = {1\over 2}\left[ (\partial \chi)^2 + (\partial\phi)^2\right] 
- \lambda(\phi^2-\lambda^2_1\chi^2)^2
\end{equation}
The ${\cal L}_{SB}$ breaks conformal invariance explicitly and was not
included in \cite{Shaposhnikov:2008a}. We shall specify it below.
We notice that ${\cal L}_M$ does not include all the terms allowed
by conformal invariance. It is possible to write down one more term
quartic in the fields, which has been set to zero. 
As explained in \cite{Shaposhnikov:2008a}, this is necessary to break 
scale invariance spontaneously.  
In the full quantum theory this is needed in order to
have a well defined perturbative expansion.
 
The model as specified above has a conformal anomaly and hence breaks scale
invariance. Within the framework of dimensional regularization this is traced
to the fact that the couplings, $\lambda$ and $\lambda_1$, are 
not dimensionless when $d\ne 4$. However
it is possible to generalize the action such that it maintains conformal
invariance in $d$ dimensions. Let us define the field $\omega$ such that
\cite{Shaposhnikov:2008a},
\begin{equation}
\omega^2 \equiv 
  (\beta^2_1 \chi^2 + \beta^2_2\phi^2)
\label{eq:omega}
\end{equation}
In $d=4-\epsilon$ dimensions, we can make all terms in the action
conformally invariant by multiplying them with a suitable power of 
$\omega$. In particular the potential term gets modified to,
\begin{equation}
(\phi^2-\lambda^2_1\chi^2)^2 \rightarrow (\phi^2-\lambda^2_1\chi^2)^2 (\omega^2)^{-\delta} 
\label{eq:potentialddim}
\end{equation}
where $\delta=(d-4)/(d-2)= -\epsilon/(2-\epsilon)$. The scalar field
kinetic energy terms as well as the term proportional to $R$ remains
unchanged.  
In $d\ne 4$, the potential terms will involve fractional powers of
the field. These terms are handled by expanding the fields around
their classical values. For example, let $\chi_0$ and $\phi_0$ 
represent the classical values of the fields $\chi$ and $\phi$ respectively
 and $\hat \chi$ and $\hat\phi$ represent
the corresponding quantum fluctuations around the classical solution. 
Hence we can express $\chi$ and $\phi$ as,
\begin{eqnarray}
\chi &=& \chi_0 + \hat\chi\nonumber\\
\phi &=& \phi_0 + \hat\phi
\label{eq:chiexpansion}
\end{eqnarray}
As long as $\chi_0\ne 0$ and $\phi_0\ne 0$, 
quantum expansion is well defined. Actually we only require the classical
value, $\omega_0$, of the field, $\omega$, defined in Eq. \ref{eq:omega},
to be non-zero. Hence a
necessary condition for a consistent perturbative expansion in this
theory is that $\omega_0\ne 0$. 
This procedure is called the GR-SI prescription in 
\cite{Shaposhnikov:2008a}.

The classical values of the scalar fields, $\chi_0$ and $\phi_0$, 
generate all the dimensional parameters in the theory, such as, the
gravitational constant, the electroweak scale, the Higgs mass etc.
As we shall see later, $\phi_0=\lambda_1\chi_0$. 
Making a quantum expansion, we find that the mass terms of the scalar fields
are given by,
\begin{equation}
{\cal L}  = -4\lambda\lambda^2_1\chi_0^2(\hat\phi-{\lambda_1}\hat \chi)^2 
\end{equation}
Hence the field, $(\hat\phi-{\lambda_1}\hat \chi)$ becomes massive. 
We shall choose the parameter range such that 
$\lambda_1<<1$. Hence the massive field is dominantly equal to $\hat\phi$. 
The orthogonal combination, proportional to, 
$(\hat\chi+{\lambda_1}\hat \phi)$,
remains massless. This field is dominantly $\hat\chi$. 

Following the procedure described above, \cite{Shaposhnikov:2008a} show that
the standard predictions of conformal invariance are preserved by 
the theory. In particular, the theory predicts a massless dilaton, 
at all orders in the perturbation theory.  
Despite the presence of
conformal invariance, the theory does predict running coupling constant. 
At one loop, \cite{Shaposhnikov:2008a,Tavares14} also argue that  
the Higgs mass is stable under quantum corrections. In the toy 
model under consideration, 
the Higgs field is identified with the field $\phi$. 
 However it has
been argued that this problem does not really get solved since, in the
presence of the Planck scale and the electroweak scale, the theory requires
some very small parameters, which have to be fine tuned at each order
\cite{Tamarit13}.

One has to impose some constraints on the parameter values in order that
the perturbation theory remains well defined.
At all orders in perturbation theory, one has to impose a constraint
on parameters, 
such that conformal invariance is  
spontaneously broken. If this is not preserved
then the perturbation theory does not make sense. Once this condition
is imposed, the theory predicts a massless dilaton in this theory.

Another important point is that removal of all divergences might require
terms of the kind, $\phi^6/\chi^2$. Such terms are allowed by scale invariance.
Hence the perturbation theory may be more complicated in 
these theories, requiring large number of parameters \cite{Tkachov09,Jain10a}.
Due to the presence of such terms, the theory is not renormalizable. 
Hence it looses predictive power at mass scale above Planck mass.
This is not a very serious problem since the additional terms
are suppressed by Planck mass. Furthermore 
it may be related to the non-renormalizability of gravity. 
This is because the scalar fields, which might lead to
such terms, are intrinsically tied to gravity. For example,
the classical values of
the fields, $\chi$ and $\phi$, generate the gravitational constant, $G$. 
 In any case, even in the presence of such terms, the consequences
of conformal invariance remain valid.  
The theory also predicts zero cosmological constant. The reason is that
it has no dimensional parameter. Hence the effective potential, at any
order in perturbation theory contains terms which are quartic in fields,
multiplied by a function of the ratio of the fields, such as $r=\phi/\chi$. 
We may express the effective potential as, $V_{eff} = 
\chi^4 U(r)$ \cite{Shaposhnikov:2008a}. 
We shall assume that $V_{eff}$ is such that, in the absence of symmetry
breaking terms, it is minimized for $\chi_0$ and $\phi_0$ not equal to
$0$ or $\pm\infty$. The minimization conditions then imply,
\begin{eqnarray}  
{dU(r)\over dr}\Bigg|_{r=r_0} &=& 0\nonumber\\
U(r_{0}) &=& 0
\label{eq:Veffmin}
\end{eqnarray}  
where $r_0={\phi_0\over\chi_0}$. 
The one loop effective potential has been explicitly constructed 
in \cite{Shaposhnikov:2008a}. 
After imposing the conditions, Eq. \ref{eq:Veffmin},  
it is found that conformal symmetry is
spontaneously broken at this order also. 
By conformal invariance, the 
 potential displays degenerate minima, such that $\chi_0$ and
$\phi_0$ take a continuous range of values and $r_0=\chi_0/\phi_0$
remains fixed. 
Eq. \ref{eq:Veffmin} also implies that,
\begin{equation}
V_{eff}(\chi_0,\phi_0) = 0
\end{equation} 
and hence leads to zero cosmological constant.

One may be concerned that the constraints, Eq. \ref{eq:Veffmin}, 
might themselves 
require fine
tuning \cite{Tamarit13} of parameters. However this does not arise, in
the following sense. Consider the potential of the model,
\begin{equation}
V(\phi,\chi) = \lambda(\phi^2-\lambda_1^2\chi^2)^2 + \lambda_2\chi^4 
\end{equation}
where we have included all the possible terms that can arise in
the potential, consistent with conformal invariance.  
The minimization conditions, Eq. \ref{eq:Veffmin}, can be satisfied only if,
\begin{equation}
\lambda_2 = 0
\label{eq:lambda2}
\end{equation}
This is not fine tuning in the sense that we do not need to maintain
a very small value of $\lambda_2$. We may compare this with the standard
problem of fine tuning of the cosmological constant \cite{Weinberg,Padmanabhan}.
The problem is most severe if we have to fine tune the cosmological
constant at each
order to a very small value. 
If we can set the cosmological constant identically to zero, even
if there is no symmetry demanding this, then this is not as severe a 
problem. Of course, ideally it would be elegant if a symmetry or some
other mechanism may demand a vanishing cosmological constant. However
in the absence of such a mechanism, it would still represent progress
if at each order in perturbation theory we don't need to fine tune 
the cosmological constant to a very small value and can simply set it to
zero. In the present case
also no symmetry requires Eq. \ref{eq:lambda2}. However we do need to impose
this constraint 
in order that perturbation theory is well
defined. Furthermore, it is satisfying that we do not need to fine tune
$\lambda_2$ to a very small value.  

We point out that there is currently considerable effort to study the
potential implications of conformal invariance in cosmology or 
high energy physics. Several model being studied are based on local
conformal invariance \cite{deser,paddy,cheng,nishino,rajpoot3,Codello,GK,quiros,bars1,Padilla}. 
The implications of global scale invariance has also been investigated
\cite{Wetterich,Armillis,Henz,Gorbanov1,Gretsch,Khoze}. 

\section{Explicit conformal symmetry breaking}
We next add a small conformal symmetry breaking term in the action.
This term is of the form, 
\begin{equation}
{\cal L}_{SB} = -{1\over 2} m_1^2\chi^2 - {1\over 2}m_2^2\phi^2 - \Lambda
+ ...
\end{equation}
Here $m_1$ and $m_2$ are the mass terms of the two fields and $\Lambda$ 
a cosmological constant. For aesthetic reasons, we may choose to set
$\Lambda=0$, but the theory does not require it.
As long as these terms are zero, such terms cannot be generated, at any order
in perturbation theory, by the action which is symmetric under conformal 
transformations. Hence 
we can choose $\Lambda$, $m_1$ and $m_2$ to be arbitrarily small 
without any fine tuning. Let us first set $\Lambda=0$. It effect will be 
discussed later.   
The basic point 
is that the mass terms lift the degeneracy in the potential.
The location of the global minimum depends on the choice of symmetry
breaking terms. We point out that by a suitable choice of such terms,
the global minimum might arise at non-zero values of $\chi$ and $\phi$.  
 At any particular
time the fields may take values such that the potential is not at its
minimum. Hence it will produce an effective cosmological constant. The
fields will also evolve slowly, as assumed in several models of inflation
\cite{Gorbanov}
or dark energy \cite{Copeland}. 
The slow roll is now controlled by the small symmetry
breaking part of the action.

We display this mechanism by a choosing simple model. We set $m_2\approx 0$. 
We choose the parameters $\beta_1$ and $\beta_2$ to be small compared to
unity. These parameters need not be very small and hence do not
require acute fine tuning. 
In the absence of symmetry breaking terms, the potential is minimized for 
\begin{equation}
\phi_0 = {\lambda_1} \chi_0  
\label{eq:minimum}
\end{equation}
We shall assume that $\lambda_1<<1$ and hence $\phi_0<<\chi_0 $. 
As we shall see, we  require $\chi_0>> M_{PL}$,
such that $\beta_1\chi_0\approx M_{PL}$. 
The value of $\lambda_1$ need not be very small, since $\phi_0$
may be of order Planck or GUT scale. Alternatively, $\phi_0$ might
be of the order of electroweak scale. In this case $\lambda_1$ does require
acute fine tuning, which is related to the standard problem of
maintaining a low Higgs mass in the presence of a very large mass scale
in the scalar potential. 
This problem is not solved in this
model  \cite{Shaposhnikov:2008a,Tamarit13}. 
In the full theory, including symmetry breaking terms, Eq. \ref{eq:minimum}
will not yield the true minimum of the potential. 

Let us first work directly in the Jordon frame and, for simplicity, just 
ignore the term proportional to $R$. As we shall, for our choice of parameters,
we get the same result in Einstein frame. We are interested in solving the 
scalar field equations of motion in order to determine the effective 
cosmological constant. The equations of motion, ignoring space derivatives 
are given by,
\begin{eqnarray}
\ddot \phi + 3H\dot \phi + 4\lambda\phi(\phi^2-\lambda^2_1\chi^2) &=& 0 \nonumber\\
\ddot \chi + 3H\dot \chi - 4\lambda\lambda^2_1\chi(\phi^2-\lambda^2_1\chi^2)
+ m_1^2\chi &=& 0
\end{eqnarray}
where $H$ is the Hubble parameter.
Assuming an approximate solution of the form, Eq. \ref{eq:minimum}, 
we find that
$\dot\phi\approx 0$.  
Here we set the second derivatives of the fields equal
to zero, since they likely to be more suppressed in comparison to the
first derivatives. We also find that, 
\begin{equation}
\dot\chi \sim m_1^2 {\chi_0 \over H}
\label{eq:dotchi}
\end{equation}
For slow roll conditions to be satisfied, we require
\begin{equation}
\dot\chi^2 << m_1^2 \chi_0^2 
\label{eq:slowroll1}
\end{equation}
which implies that,
\begin{equation}
m_1<< H
\end{equation}
Hence for slow roll, we require that the symmetry breaking terms are much 
smaller than the Hubble parameter. Such small terms would normally require
acute fine tuning. However in the present case these are protected by
conformal symmetry.

The solution leads to vacuum energy equal to $m_1^2\chi_0^2/2$. 
Hence, in order that it generates a sufficiently large value of the effective
cosmological constant, we require,
\begin{equation}
 m_1^2 \chi_0^2\sim M_{PL}^2 H^2 
\label{eq:cosmic evolution}
\end{equation}
In the Jordon frame, the gravitational constant undergoes a slow evolution,
which has been ignored in the above equation. This evolution can be
consistently ignored as long as the slow roll conditions are satisfied.
It is useful to perform the entire calculation in the Einstein frame which,
as we shall see below,
leads to the same result.
Eq. \ref{eq:cosmic evolution}, along with the slow roll condition leads
to the constraint,
\begin{equation}
 \chi_0 >> M_{PL} 
\label{eq:vacuumchi}
\end{equation}

We point out that the model contains some small parameters, such as, $\beta_1$ 
 and $\lambda_1$, which are not protected by conformal invariance.
However these parameters need not be very small. Their precise values
depend on the model under consideration. For our purposes these may
be of the order of $10^{-3}$. The possibility that $\lambda_1$ may be
very small and its associated fine tuning has already been discussed above. 
The important point is that the mass parameters, such as, $m_1$,
are extremely tiny in comparison to other mass scales, such as $M_{Pl}$,
$\phi_0$ etc.  
 Their small value, however, is protected against quantum corrections
by conformal invariance. 

We next perform the calculation in the Einstein frame. We make the conformal
transformation, such that,
\begin{equation}
g_{\mu\nu} \rightarrow \Omega^2 g_{\mu\nu}
\label{eq:conformalgmunu}
\end{equation}
where 
\begin{equation}
\Omega^2 \equiv {M_{PL}^2\over \omega^2} 
\label{eq:Omega}
\end{equation}
The Lagrangian density in terms of the transformed variables can be written
as,
\begin{eqnarray}
{\cal L}_E &=& - {M_{PL}^2\over 16 \pi} R - {3M_{PL}^2\over 8 \pi} 
g^{\mu\nu}\partial_\mu\ln\Omega\partial_\nu\ln\Omega\nonumber\\
&+& {\Omega^2\over 2} g^{\mu\nu}\partial_\mu\phi \partial_\nu\phi
+ {\Omega^2\over 2} g^{\mu\nu}\partial_\mu\chi \partial_\nu\chi
-\Omega^4 V
\label{eq:Leinsteinframe}
\end{eqnarray}
where $V$ is the potential,
\begin{equation}
V = \lambda(\phi^2-\lambda^2_1\chi^2)^2 + 
{1\over 2} m_1^2\chi^2 + {1\over 2}m_2^2\phi^2
\label{eq:Potential}
\end{equation}
where, as before, we shall assume, $m_2\approx 0$. We next
obtain the equations of motion keeping only the time derivative.
 Using the slow roll approximation we drop the second derivatives. Under
this approximation, we obtain,
\begin{eqnarray}
-3H\dot \phi + {9H\over 4\pi}{\beta^2_2 \phi\over \omega^2}(\beta^2_1\chi\dot\chi
+\beta^2_2\phi\dot\phi) - {M_{PL}^2\over\omega^2} {\partial V\over \partial\phi}
+{4M_{PL}^2\over \omega^4}\beta^2_2\phi V &=& 0\nonumber\\
-3H\dot \chi + {9H\over 4\pi}{\beta^2_1 \chi\over \omega^2}(\beta^2_1\chi\dot\chi
+\beta^2_2\phi\dot\phi) - {M_{PL}^2\over\omega^2} {\partial V\over \partial\chi}
+{4M_{PL}^2\over \omega^4}\beta^2_1\chi V &=& 0
\label{eq:Einsteinframe}
\end{eqnarray}
In the absence of symmetry breaking terms we again find the same result
as Eq. \ref{eq:minimum}, with $\dot \phi=0$ and $\dot\chi=0$. 
Solving the full equations, assuming the relationship Eq. \ref{eq:minimum},
between classical values of $\phi$ and $\chi$, we find that both
$\dot\phi$ and $\dot\chi$ are related to the symmetry breaking terms. 
The second terms on the left hand side of both the equations are negligible
since, $\beta_1<<1$ and $\beta_2<<1$. Given that, at leading order, 
$\omega\sim \beta_1\chi_0
\sim M_{PL}$, we again find that $\dot\chi$ is given by Eq. \ref{eq:dotchi}.  
In the present case $\dot\phi\ne 0$. However
it is clear that $\dot\phi<< \dot\chi$, being suppressed by the factor
$\beta^2_2\phi_0/(\beta_1^2\chi_0)$. 
Hence we again get exactly the same condition, Eq. \ref{eq:vacuumchi}, 
as obtained in the Jordon frame. 

\subsection{Higher Orders}
The above analysis may be performed at any order in perturbation 
theory using the effective potential. 
While computing the quantum contributions to effective potential, 
we ignore the symmetry breaking
terms. The symmetry breaking terms are assumed to be extremely small
and hence are expected to give negligible contributions at higher orders
to the effective potential.  
Hence the effective potential at any order
can be expressed as,
\begin{equation}
V_{eff} = \chi^4 U(r) + {1\over 2}m_1^2\chi^2 + {1\over 2} m_2^2\phi^2
\label{eq:VeffHigher}
\end{equation}
where, the term $\chi^4 U(r)$ is obtained entirely from the symmetry 
preserving part of the action.
We require that, in the absence of symmetry breaking terms, 
at each order the effective potential displays a minimum,
where its value is nonzero and finite. 
Hence we have to impose some conditions on
the counter terms so that this holds \cite{Shaposhnikov:2008a}. 
Due to conformal invariance this minimum value of the potential
can only be zero. 

We now replace $V$ in  Eq. \ref{eq:Einsteinframe} by $V_{eff}$. 
We are interested in a solution subject to the conditions specified by
Eq. \ref{eq:Veffmin}. Imposing these conditions in Eq. \ref{eq:Einsteinframe},
we find that, $\dot \phi$ and $\dot \chi$ are both
proportional to symmetry breaking terms. The value of $\dot\chi$ is again
given by Eq. \ref{eq:dotchi} and $\dot\phi<<\dot\chi$. 
Hence we can maintain their
small values without any fine tuning.  

\subsection{Non-zero cosmological constant}
We next discuss the case where the symmetry breaking terms contain a 
non-zero cosmological constant, $\Lambda$. In this case we set the masses,
$m_1$ and $m_2$, equal to zero. 
This case is very simple. The degeneracy of the minimum does not get 
lifted, i.e. the minimum is exactly degenerate even when we include
symmetry breaking terms. 
Hence the equations of motion satisfy Eq. \ref{eq:minimum} exactly. 
The theory now has non-zero cosmological constant. However it 
does not receive large corrections from the symmetry preserving terms. 
At higher orders also Eq. \ref{eq:minimum} 
is maintained by a suitable choice of counter terms
\cite{Shaposhnikov:2008a}. 
Hence we still have a degenerate minima, with the minimum value
approximately equal to zero, up to the corrections due to symmetry breaking
term, $\Lambda$. 
 
\section{Applications to inflation and dark energy}
The mechanism that we have discussed above may be applied either to 
inflation or to dark energy. Let us first discuss the case of inflation. 
In this case it is simplest
to choose symmetry breaking terms such that $\Lambda=0$. We can choose the
mass terms, $m_1$ and $m_2$, sufficiently small to satisfy the slow roll 
conditions. Inflation ends when the fields reach the true minima of
the potential. The phenomenon acts like the standard large field inflation
\cite{Gorbanov}. The symmetry breaking terms have to be of the order
of the inflationary scale. Hence this theory will have conformal breaking
of the order of inflationary scale and will not solve the fine tuning problem of
dark energy. However the inflationary slow roll condition can be
met without any fine tuning.  

Alternatively we may accommodate inflation by fine tuning the symmetry
preserving terms and the symmetry breaking terms may be only of
the order of dark energy. In this case we may either introduce an
explicit cosmological constant or masses, $m_1$
and $m_2$. In case of cosmological constant, the constraint
on the field $\chi$, Eq. \ref{eq:vacuumchi}, is not applicable.  
However this constraint is applicable if dark energy is generated
by the masses, which leads to a slow evolution of the fields.

\section{Conclusions}
In this paper we have shown that conformal symmetry provides a mechanism
which partially alleviates the problem of fine tuning of the cosmological
constant. We use the GR-SI prescription in which the conformal invariance
can be maintained in the full quantum theory \cite{Shaposhnikov:2008a}. 
However the perturbation
theory gets more complicated and renormalizability of the theory does not
remain maintained \cite{Tkachov09,Jain10a}. Hence the theory looses 
predictability beyond a certain mass scale, which in the present
model is taken to be the Planck scale. Hence this absence of
renormalizability is not a very serious issue at low energies.  
The conformal invariance in the theory is spontaneously broken for a
certain range of parameters. The
perturbation theory makes sense only if this can be accomplished. Hence
we have to impose an additional constraint on the theory, not required
by conformal invariance. We have argued that this constraint does not
amount to fine tuning of a parameter since it does not involve maintaining
a small value of a parameter at each order in perturbation theory.
It simply requires setting some parameter value identically to zero.
Given this constraint, the perturbation theory can be well defined.

If we impose exact conformal invariance on the theory, then it predicts
zero cosmological constant. We introduce small conformal symmetry breaking
terms. These involve mass terms of scalar fields and/or explicit 
cosmological constant. Since the symmetry preserving part does not 
generate such terms at any order in the perturbation theory, we can
maintain their small value without any fine tuning. 
We may identify the cosmological constant with dark energy. Alternatively
the scalar mass terms lead to slowly rolling scalar field and hence
can also generate dark energy. Another possibility is that the model may
be applied to generate inflation. Detailed application of the
model to dark energy or inflation is not pursued in this paper.

\bigskip
\noindent
{\bf \large Acknowledgements:}  Gopal Kashyap thanks the Council
of Scientific and Industrial Research (CSIR), India for providing his Ph.D.
fellowship. We thank Joydeep Chakrabortty for useful discussions.

\end{spacing}
\begin{spacing}{1}
\begin{small}

\end{small}
\end{spacing}
\end{document}